\documentclass[font size=10pt]{scrartcl}
\usepackage[right=2cm, left=2cm]{geometry}

\usepackage{hyperref}
\usepackage[english]{babel}
\usepackage[utf8]{inputenc}
\usepackage{color}
\usepackage{cite}
\usepackage{mathtools}
\usepackage{caption}         
\usepackage{physics}
\usepackage{graphicx}

\title{\raggedright Hybrid static potentials in SU(2) lattice gauge theory at short quark-antiquark separations}
\author{ Carolin Riehl\thanks{Speaker} , Marc Wagner \\
	\normalsize Goethe-Universit\"at Frankfurt, Institut f\"ur Theoretische Physik, Max-von-Laue-Stra{\ss}e 1, \vspace{-0.2cm} \\ \normalsize D-60438 Frankfurt am Main, Germany \\
	\normalsize Helmholtz Research Academy Hesse for FAIR, Campus Riedberg, Max-von-Laue-Stra{\ss}e 12, \vspace{-0.2cm} \\ \normalsize D-60438 Frankfurt am Main, Germany}
\date{}
\publishers{\vspace{0.4cm} \normalsize Asia-Pacific Symposium for Lattice Field Theory -- APLAT 2020 \\
	\normalsize 4-7 August 2020}

\newcommand{\gtapprox}{\raisebox{-0.5ex}{$\,\stackrel{>}{\scriptstyle\sim}\,$}}

\newcommand{\RN}[1]{\uppercase\expandafter{\romannumeral#1}}

\begin{document}
	
	\maketitle
	
	\begin{abstract}{ \textbf{Abstract.}}
		We compute hybrid static potentials in SU(2) lattice gauge theory using a multilevel algorithm and three different small lattice spacings. The resulting static potentials, which are valid for quark-antiquark separations as small as $0.05\, \text{fm}$, are important e.\,g.\ when computing masses of heavy hybrid mesons in the Born-Oppenheimer approximation. We also discuss and exclude possible systematic errors from topological freezing, the finite lattice volume and glueball decays.
	\end{abstract}
	
	\fontsize{12pt}{14pt}\selectfont
	
	\section{Introduction}
	Hybrid static potentials represent the energy of an excited gluon field surrounding a static quark and a static antiquark as a function of their separation and are, thus, related to heavy hybrid mesons.
	Due to the gluonic excitations, quantum numbers of hybrid mesons can be different from those predicted by the constituent quark model.
	The investigation of exotic mesons like hybrid mesons and tetraquarks are currently a very active field of research, both theoretically and experimentally (for reviews cf.\ e.\,g.~\cite{Braaten:2014ita,Meyer:2015eta,Swanson:2015wgq,Lebed:2016hpi,Olsen:2017bmm,Brambilla:2019esw}).
	
	One possibility to compute masses of heavy hybrid mesons is the Born-Oppenheimer approximation~\cite{Perantonis:1990dy,Juge:1997nc,Juge:1999ie,Braaten:2014qka,Capitani:2018rox}. 
	In a first step, the heavy quark and the heavy antiquark are considered as static and lattice gauge theory is used to compute the energy of the gluons. 
	The resulting hybrid static potentials are then parameterized by analytic functions. 
	In a second step, these potentials are inserted into the Schr\"odinger equation for the relative coordinate of the heavy quark-antiquark pair. Solving the Schr\"odinger equation leads to energy eigenvalues, which can be interpreted as masses of heavy hybrid mesons.
	Besides their importance for the computation of heavy hybrid meson masses, hybrid static potentials are also relevant in the context of effective field theories like potential Non-Relativistic QCD (pNRQCD), where their short-distance behavior fixes the matching coefficients~\cite{Berwein:2015vca,Oncala:2017hop,Brambilla:2018pyn}. 
	
	In this work, we compute hybrid static potentials in $SU(2)$ lattice gauge theory at three different lattice spacings, $a = 0.026 \, \text{fm}$\,, $0.041 \, \text{fm} $\,, $0.077 \, \text{fm}$\,, significantly smaller than those used in previous works, e.\,g.\ in Refs.~\cite{Juge:2002br,Bali:2003jq,Capitani:2018rox}.
	In particular, we present results for the ordinary and the two lowest hybrid static potentials for quark-antiquark separations as small as $0.05 \, \text{fm}$.
	We also discuss and exclude possible systematic errors from topological freezing, the finite lattice volume and glueball decays.
	
	\section{Hybrid static potentials}
	Quantum numbers of (hybrid) static potentials are the following:
	\begin{itemize}
		\item $\Lambda = \Sigma (=0), \Pi (=1), \Delta (=2), \dots $ denotes non-negative integer values of the total angular momentum with respect to the quark-antiquark separation axis. 
		\item $\eta= g,u$ describes the even ($g$) or odd ($u$) behavior under the combined parity transformation and charge conjugation, $\mathcal{P}\circ \mathcal{C}$.
		\item $\epsilon=+,-$ is the eigenvalue of a reflection $\mathcal{P}_x$ along an axis perpendicular to the quark-antiquark separation axis. 
		Hybrid static potentials with $\Lambda \ge 1$ are degenerate with respect to $\epsilon$.
	\end{itemize}
	The ordinary static potential is labeled by $\Lambda_{\eta}^{\epsilon}=\Sigma_g^+$, while the two lowest hybrid static potentials have quantum numbers $\Pi_u$ and $\Sigma_u^-$.
	
	Hybrid static potentials are computed from Wilson loop-like correlation functions on $SU(2)$ gauge link configurations.
	To excite gluons with quantum numbers different from that of the ordinary static potential, we employ creation operators, where non-trivial shapes replace the straight spatial parallel transporters. 
	More precisely, we use the creation operators  $S_{\RN{3},1}$ and $S_{\RN{4},2}$ defined in Table $3$ and Table $5$ in Ref.~\cite{Capitani:2018rox}, where the creation operators are discussed in detail.

	\section{Numerical results}
	\subsection{Lattice setup}
	All computations were performed on $SU(2)$ gauge link configurations, which were generated with a Monte Carlo heatbath algorithm with the Wilson plaquette action.
	Additionally, we used a multilevel algorithm~\cite{Luscher:2001up}, which leads to an exponential error reduction in the expectation values of the Wilson loop-like observables.
	We observed that one level is sufficient, where we split the lattice into timeslices of thickness $2a$, which were updated more often than the full lattice.
	The expectation values of the Wilson loop-like observables were then computed from products of timeslice averages of two-link operators.
	
	We generated three ensembles with different values of the gauge coupling, $\beta=2.50$\,, $2.70$\, and $2.85$\,, which correspond to lattice spacings $a = 0.077 \, \text{fm}$\,, $0.041 \, \text{fm}$\, and $0.026 \, \text{fm}$\,, respectively.
	This scale setting is taken from Ref.\ \cite{Hirakida:2018uoy} and based on the gradient flow, the $t_0$ scale and the identification $r_0 = 0.5 \, \text{fm}$\,, where $r_0$ is the Sommer scale.
	
	Hybrid static potentials were extracted from plateaus of the effective potentials.
	The contributions of excited states to the effective potentials were reduced to a minimum by employing the creation operator shapes, which were optimized at a lattice spacing $a = 0.093 \, \text{fm}$\, in Ref.~\cite{Capitani:2018rox} and keeping the operator extents in physical units constant.
	Moreover, we applied APE smearing~\cite{Albanese:1987ds} on the spatial links with $\alpha_{\text{APE}}=0.5$. 
	The number of APE smearing steps was optimized for each lattice spacing, i.\,e.\ $N_{\text{APE}}$ had to be increased with decreasing lattice spacing.
	
	\subsection{Hybrid static potentials}
\begin{figure}\centering
		\includegraphics[width=0.8\linewidth]{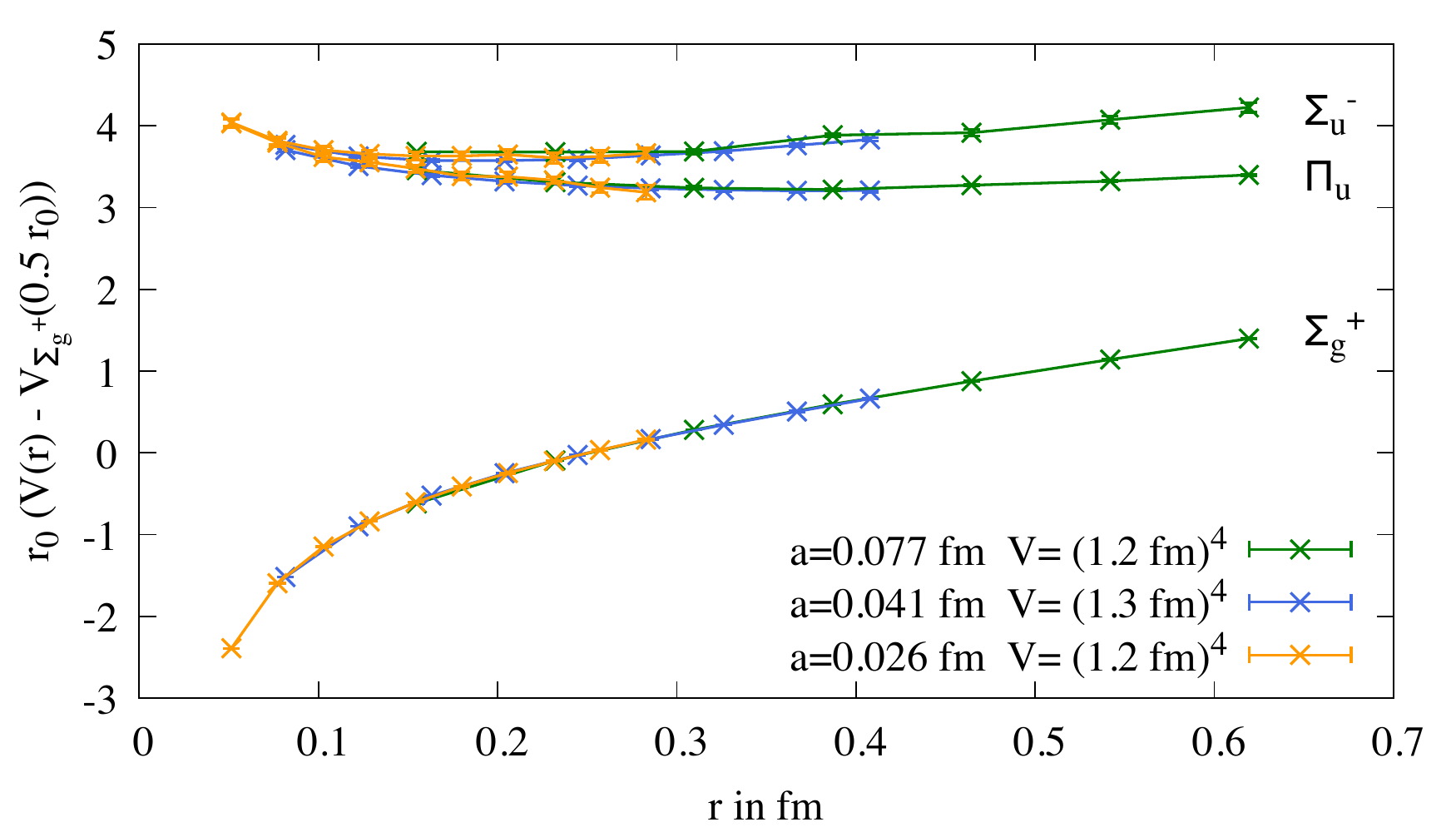}
		\caption{Lattice results for the ordinary static potential $\Sigma_g^+$ and the hybrid static potentials $\Pi_u$ and $\Sigma_u^-$.}
		\label{FIG1}
	\end{figure}

	In Figure~\ref{FIG1} we show our lattice results for the ordinary static potential $\Sigma_g^+$ and the two lowest hybrid static potentials $\Pi_u$ and $\Sigma_u^-$ for three different lattice spacings as functions of the quark-antiquark separation $r$. For each potential, data points with $r \geq 2 a$ are consistent with a single curve. We interpret this as an indication that lattice discretization errors for $r \geq 2 a$ are negligible.
	
	Thus we are able to present lattice results for hybrid static potentials (for $SU(2)$) for separations as small as $r \approx 0.05 \, \text{fm}$\,, while previous lattice results for hybrid static potentials (for $SU(3)$) were provided and trustworthy only for separations $r \gtapprox 0.16 \,\text{fm}$~\cite{Juge:2002br,Bali:2003jq,Capitani:2018rox}.
	We plan to extend our computation of hybrid static potentials at small lattice spacings and small quark-antiquark separations to $SU(3)$ in the near future and use the corresponding results for the computation of heavy hybrid meson masses in the Born-Oppenheimer approximation.
	
	\section{Excluding possible systematic errors}
	\subsection{\label{SEC001}Finite spatial lattice volume}
	When performing computations at different spatial lattice volumes we observed a sizable volume dependence of the ordinary static potential and hybrid static potentials, when the volume is smaller than $\approx  (1.0 \,\text{fm})^3$.
	For example, shrinking the spatial lattice volume causes a small negative shift for the ordinary static potential $\Sigma_g^+$, while there is a much larger positive shift for the hybrid static potential $\Pi_u$.
	
	For a spatial lattice volume of $(1.2 \,\text{fm})^3$, however, as e.\,g.\ used for the computation of the potentials shown in Figure~\ref{FIG1}, these finite volume corrections are already tiny and negligible compared to statistical errors. 
	
	\subsection{Topological freezing}
	Gauge field configurations can be classified according to their topological charge.
	\textit{Topological freezing} denotes the problem that a Monte Carlo simulation is trapped in one of the topological sectors.
	The gauge link configurations generated in such a simulation do not form a representative set distributed according to $e^{-S}$.
	This problem might appear, when using a lattice spacing $a$ smaller than $\approx 0.05 \, \text{fm}$ \cite{Luscher:2011kk} and becomes more severe, when approaching the continuum limit, i.\,e.\ when further decreasing $a$.
	If a simulation is trapped in a topological sector, observables exhibit specific finite volume corrections in addition to those discussed in section~\ref{SEC001} (see e.\,g.\ Refs.~\cite{Brower:2003yx,Aoki:2007ka,Bietenholz:2016ymo}).
	
\begin{figure}\centering
		\includegraphics[width=0.6\linewidth]{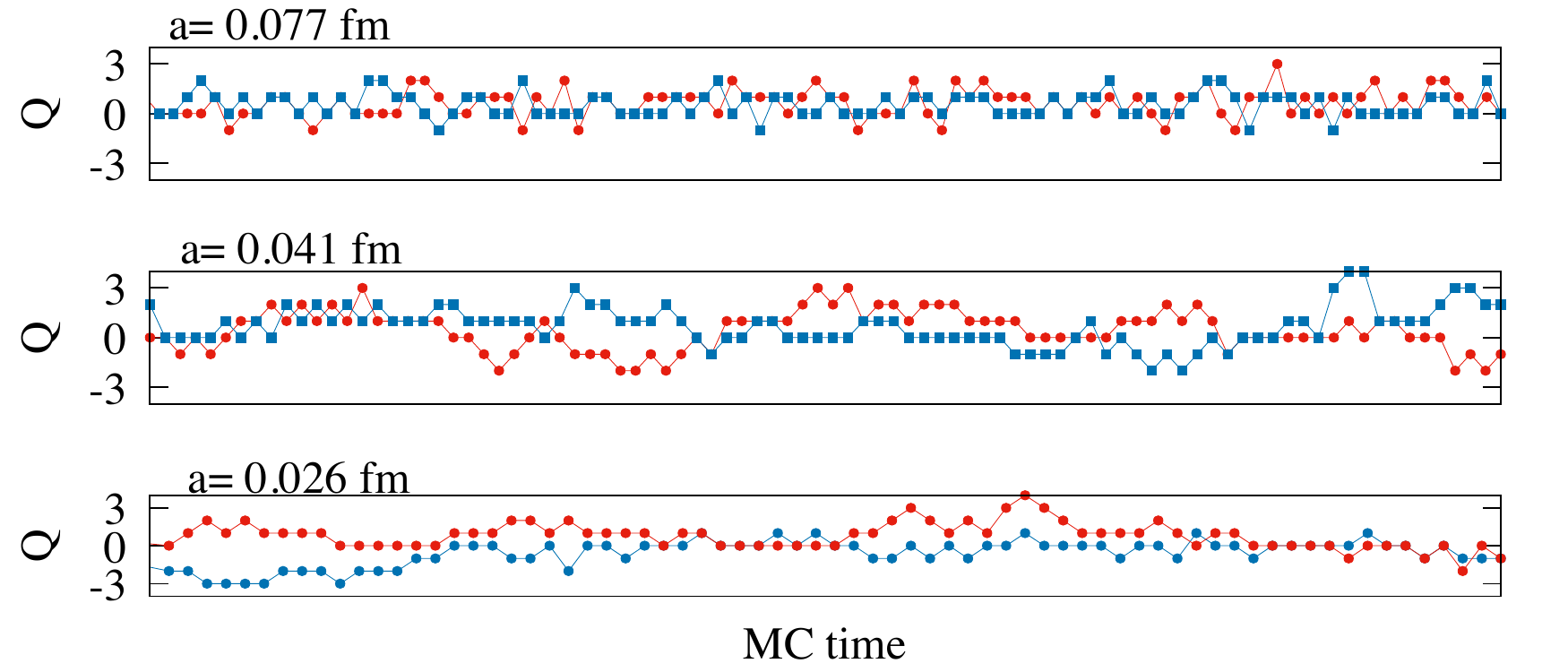}
		\caption{Topological charge as a function of the Monte Carlo time for two independent exemplary runs for each lattice spacing. Note that the horizontal axis corresponds to only small sections ($\approx 25 \%$)  of the full Monte Carlo histories of these runs.}
		\label{FIG2}
	\end{figure}

	To check, whether our simulations suffer from topological freezing, we computed the topological charge on each gauge link configuration via a field-theoretic definition with a simple clover-leaf discretization and 4-dimensional APE-smearing \cite{Cichy:2014qta}. 
	From Figure~\ref{FIG2} one can see that the topological charge still changes frequently at all our lattice spacings.
	The topological charge distribution and topological susceptibility (which we will show and discuss in a future more detailed publication) also indicate that the Monte Carlo algorithm is able to sample the gauge link configurations correctly.
	Furthermore, through a suitable binning and several independent Monte Carlo runs we exclude that statistical errors are underestimated, because of autocorrelations, which are also expected to increase with decreasing lattice spacing.
	Thus, the potentials presented in this work should be free of any systematic errors from topological freezing.
	
	\subsection{Glueball decay}	
	For sufficiently small quark-antiquark separations $r$ the energy difference between a hybrid static potential and the ordinary static potential $\Sigma_g^+$ is large enough such that the hybrid flux tube can dissolve into a glueball and the $\Sigma_g^+$ flux tube.
	The minimal energy, which is necessary for a decay into the lightest glueball with quantum numbers $J^{PC}=0^{++}$ and mass $m_{0^{++}}=4.21/r_0$\cite{Morningstar:1999rf} is shown as a dashed line in Figure~\ref{FIG3} together with our previous lattice results for hybrid static potentials from Ref.\ \cite{Capitani:2018rox}.
	
	\begin{table}[t]\centering
		\begin{minipage}{\linewidth}\centering
			\includegraphics[width=0.75\linewidth]{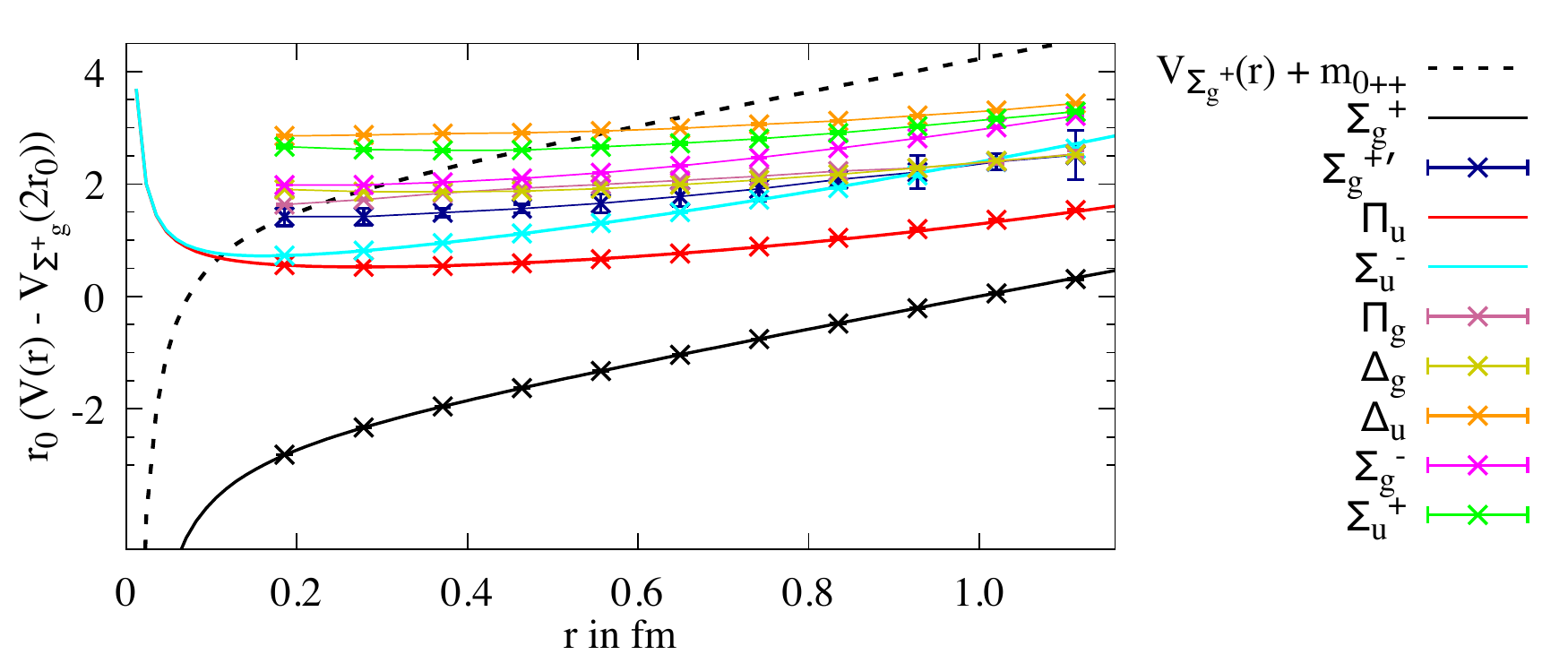}
			\captionof{figure}{Threshold energy $V_{\Sigma_g^+}(r) + m_{0^{++}}$ for the decay of a hybrid flux tube to the flux tube of the ordinary static potential and the lightest $0^{++}$ glueball. Lattice results for static potentials are taken from Ref.\ \cite{Capitani:2018rox}.}
			\label{FIG3}
		\end{minipage}
		\begin{minipage}{\linewidth}\centering
			\begin{tabular}{|c|cccc|cc|cc|}\hline
				$\Lambda_{\eta}^{\epsilon}$ & $\Pi_u$    & $\Pi_g$    & $\Delta_g$ & $\Delta_u$ & ${{\Sigma}_g^+}^{\prime}$ & $\Sigma_u^+$ & $\Sigma_u^-$ & $\Sigma_g^-$\\ \hline
				$r_{\text{crit}}$ in fm         & $0.11$ & $0.23 $ & $0.28 $ & $0.58 $ &$0.19 $  &$0.46 $  &$0.06 $  & $0.12 $ \\\hline
			\end{tabular}
			\caption{Approximate separations $r_{\text{crit}}$, below which decays of hybrid flux tubes to the flux tube of the ordinary static potential and the lightest $0^{++}$ glueball (or in the case of $\Sigma_u^-$ and $\Sigma_g^-$ the next lightest $2^{++}$ glueball)  become energetically possible. }
			\label{TAB1}
		\end{minipage}
	\end{table}
	
	From this plot, we can estimate separations $r_{\text{crit}}$, below which a glueball decay is energetically allowed.
	These separations are listed in Table~\ref{TAB1}.
	For $r < r_{\text{crit}}$ a hybrid static potential creation operator might generate non-vanishing overlap to the $\Sigma_g^+$ flux tube and a glueball.
	In Figure~\ref{FIG1} we show lattice results for the hybrid static potentials $\Pi_u$ and $\Sigma_u^-$ for separations below $r_{\text{crit}} \approx 0.1 \, \text{fm}$.
	There is, however, no sign of contamination by glueball decay.
	Indeed, the two lowest hybrid static potentials reveal the expected increasing behavior and degeneracy at small separations \cite{Berwein:2015vca}.
	
	For the $\Sigma_u^-$ and $\Sigma_g^-$ potentials it is even possible to exclude decays to the lightest $0^{++}$ glueball using symmetry arguments (we will discuss this in detail in a future publication). 
	Still allowed are decays into the next lightest glueball with quantum numbers $J^{PC}=2^{++}$. 
	However, this is energetically only possible for very small separations.
	

	\section*{Acknowledgments}
	We thank Christian Reisinger for providing his multilevel code and helpful conversations. 
	We acknowledge useful discussions with Colin Morningstar and Joan Soto.
	
	M.W.\ acknowledges support by the Heisenberg Programme of the Deutsche Forschungsgemeinschaft (DFG, German Research Foundation) -- project number 399217702.
	
	Calculations on the GOETHE-HLR and on the FUCHS-CSC high-performance computers of the Frankfurt University were conducted for this research. We would like to thank HPC-Hessen, funded by the State Ministry of Higher Education, Research and the Arts, for programming advice.


\end{document}